\documentclass[colorlinks,balance,nofoot,nohead]{asmeconf}

\usepackage{multirow}
\usepackage{stfloats}
\usepackage{microtype}
\newcolumntype{x}[1]{>{\raggedright\arraybackslash}p{#1}}

\newcommand{\openj}{\textsc{OpenJ}}
\newcommand{\jack}{Tecnomatix Jack}

\definecolor{scholarblue}{HTML}{1A0DAB}

\hypersetup{%
	pdfauthor={Bank, Eaton},
	pdftitle={OpenJ: A Conceptual Framework for Open-Source Digital Human Modeling and Ergonomic Assessment in a CAD Environment},
	pdfkeywords={Digital Human Modeling, Biomechanics, Human Performance/Force Assessment, Computer-Aided Design, Software},
	pdfsubject={arXiv preprint -- IMECE 2026},
	linkcolor=scholarblue,
	citecolor=scholarblue,
	urlcolor=scholarblue,
	filecolor=scholarblue,
}

\begin{document}


\title{\textsc{OpenJ}: A Conceptual Framework for Open-Source Digital Human Modeling and Ergonomic Assessment in a CAD Environment}

\SetAuthors{%
    H.~Sinan~Bank\affil{1}\CorrespondingAuthor{sinan.bank@colostate.edu},
    Casey~E.~Eaton\affil{2}
}

\SetAffiliation{1}{Department of Systems Engineering, Colorado State University, Fort Collins, CO 80523}
\SetAffiliation{2}{Department of Industrial and Systems Engineering, Auburn University, Auburn, AL 36849}

\microtypesetup{nopatch=item}
\maketitle
\microtypesetup{patch=item}

\keywords{Digital Human Modeling, Biomechanics, Human Performance/Force Assessment, Computer-Aided Design, Software}


\begin{abstract}
Industrial workplace challenges range from musculoskeletal disorders---a leading cause of occupational injury---to suboptimal workstation layouts, inefficient task sequences, and poor human-equipment fit. Digital human modeling (DHM) tools address several of these challenges by placing a scalable virtual mannequin in a computer-aided design (CAD) environment, enabling engineers to evaluate ergonomic risk through standardized assessment methods (RULA, REBA, NIOSH Lifting Equation, OWAS), optimize workstation layouts for reach and visibility, predict task postures through inverse kinematics, and simulate operations before physical implementation. Despite four decades of development since the Jack system originated at the University of Pennsylvania in the 1980s, the integrated DHM capability set---anthropometric mannequin, posture prediction, ergonomic assessment, and CAD integration---remains exclusive to commercial platforms such as Siemens Tecnomatix Jack (Process Simulate), Dassault DELMIA, Humanetics RAMSIS, and the University of Iowa's Santos system. These platforms operate under proprietary, vendor-quoted pricing models, and their acquisition and operating costs together with closed-source implementations have been repeatedly identified as practical adoption barriers for individual researchers, small-to-medium enterprises, and educational institutions. Organizations without access resort to manual observational methods---paper-based worksheets applied to photographs or video---sacrificing the predictive power and reproducibility that computational analysis provides.

Meanwhile, the open-source ecosystem offers capable but disconnected components across adjacent domains: musculoskeletal simulators (OpenSim, biorbd), parametric body generators (MakeHuman), physics engines (MuJoCo, PyBullet), markerless motion capture pipelines (Pose2Sim), and scattered ergonomic calculators. A systematic search across open-source repositories, peer-reviewed biomechanics software publications, and the broader academic literature confirms that no existing tool integrates a scalable anthropometric mannequin, standard ergonomic assessment methods, posture prediction, and CAD environment integration into a single workflow.

This paper presents the conceptual framework for \openj{}, an open-source digital human modeling and ergonomic assessment tool designed to bridge this gap. The framework is grounded in a review of 52~references spanning eight research domains: commercial DHM, open-source biomechanics tools, anthropometric databases (ANSUR~II), posture prediction methods, ergonomic assessment algorithms, reach and vision analysis, task simulation, and FreeCAD workbench architecture. Five design objectives are identified, tracing from the unmet integration gap through twelve system functions. The proposed two-layer architecture will be built around a standalone Python core library (\texttt{openj-core})---usable headlessly from scripts and Jupyter notebooks, and the primary interface for batch analyses, CI pipelines, and integration into external workflows---with a FreeCAD workbench (\texttt{openj\_wb}) as one frontend included for visual and educational use; this follows the core-library-plus-Python-bindings pattern used by biorbd. This separation will enable the core library to be tested independently and extended with future frontends without modifying the computational engine.

The framework addresses the needs of three underserved stakeholder groups: ergonomics researchers requiring modifiable and reproducible analyses with a scriptable Python API, industrial engineers at small-to-medium enterprises who need zero-cost ergonomic evaluation tools, and engineering educators seeking hands-on DHM instruction capabilities for their courses. Core technical components include a three-tier anthropometric scaling pipeline from ANSUR~II population data with de~Leva body segment parameters, optimization-based posture prediction with comfort weighting complemented by Pinocchio-based differential inverse kinematics for interactive-speed manipulation, a pluggable assessment architecture accommodating methods with varying automation levels (five built-in plugins: RULA, REBA, NIOSH, OWAS, and Snook tables), Monte Carlo reach envelope computation, asymmetric vision cone modeling with ray-casting occlusion, and a lightweight task simulation engine for cumulative exposure analysis. All major design decisions are traced to peer-reviewed literature, and a validation plan with quantitative acceptance criteria referencing published scoring examples is outlined. The paper serves as a design blueprint for \openj{} (Open-Jane/Joe), positioning the project for subsequent open-source implementation and community adoption.
\end{abstract}


\begin{nomenclature}
\EntryHeading{Acronyms}
\entry{ANSUR}{Anthropometric Survey of U.S.~Army Personnel}
\entry{BSP}{Body Segment Parameters}
\entry{CAD}{Computer-Aided Design}
\entry{DHM}{Digital Human Modeling}
\entry{DOF}{Degrees of Freedom}
\entry{IK}{Inverse Kinematics}
\entry{NIOSH}{National Institute for Occupational Safety and Health}
\entry{OWAS}{Ovako Working Posture Analysing System}
\entry{REBA}{Rapid Entire Body Assessment}
\entry{RULA}{Rapid Upper Limb Assessment}
\entry{URDF}{Unified Robot Description Format}
\end{nomenclature}


\section{Introduction}\label{sec:introduction}

Industrial workplace challenges range from musculoskeletal disorders (MSDs)---which accounted for roughly one-third of nonfatal occupational injury and illness cases involving days away from work in 2015~\cite{BLS2016} and remained a leading event category in the most recent U.S.\ Bureau of Labor Statistics summary, with over half a million reported MSD cases involving days away from work across 2021--2022~\cite{BLS2024}---to suboptimal workstation design, inefficient task sequencing, and poor human-equipment fit. Digital human modeling (DHM) tools address several of these challenges simultaneously: by placing a biomechanically representative virtual mannequin inside a computer-aided design (CAD) environment, engineers can evaluate ergonomic risks, optimize workspace layouts, and validate operational sequences before physical prototypes are built. The Jack system, developed at the University of Pennsylvania in the mid-1980s under Norman~I.~Badler with NASA and U.S.~Army funding~\cite{Badler1993,Blanchonette2010}, introduced the integrated capability set---parametric mannequin, inverse-kinematics posture prediction, standardized ergonomic scoring, and reach/vision analysis---that subsequent commercial DHM platforms have adopted.

Despite four decades of maturation, this integrated capability set remains locked behind commercial licenses. Siemens \jack{}, Dassault DELMIA, Humanetics RAMSIS, and the University of Iowa's Santos~\cite{Chaffin2001,Chaffin2005,AbdelMalek2007,VanDerMeulen2007} collectively serve the market under proprietary, vendor-quoted pricing models whose acquisition and ongoing costs---together with associated training and integration overhead---have been repeatedly identified as practical adoption barriers in occupational safety and ergonomics workflows~\cite{Schall2018,GonzalezAlonso2024}. Comparative studies have also documented analytical discrepancies between platforms: Pol\'{a}\v{s}ek et al.~\cite{Polasek2015} reported approximately 13\% differences in NIOSH Recommended Weight Limit (RWL) values between Jack and DELMIA for identical scenarios, attributing the difference to the greater granularity of user-exposed input parameters in Jack (recovery time, uninterrupted work time, and grip detail).

Meanwhile, the open-source ecosystem offers mature components in adjacent domains---musculoskeletal simulation (OpenSim~\cite{Delp2007,Seth2018}, biorbd~\cite{Michaud2021}), parametric body generation (MakeHuman~\cite{Paul2019}), rigid-body dynamics (MuJoCo~\cite{Todorov2012}), and markerless motion capture (Pose2Sim~\cite{Pagnon2022})---but these remain disconnected fragments. A survey of repositories, software publications, and academic databases confirms that \emph{no existing fully open-source tool integrates all four capabilities---a scalable anthropometric mannequin, posture prediction, ergonomic assessment, and CAD environment---into a single workflow}~\cite{Michaud2021,Pagnon2022,Skuric2023}.

This paper presents the conceptual framework for \openj{}, an open-source DHM tool designed to fill this gap. \openj{} proposes a two-layer architecture in which a standalone Python core library (\texttt{openj-core}) will carry the computational engine and be usable headlessly, while a FreeCAD workbench frontend (\texttt{openj\_wb}) will provide one visual interface among possible future ones; this follows the core-library-plus-Python-bindings pattern of biorbd~\cite{Michaud2021}. Five design objectives are decomposed into twelve system functions, and five core technical components are specified with traceability to peer-reviewed literature.

The remainder of the paper is organized as follows. Section~\ref{sec:background} surveys the commercial and open-source landscapes and identifies the integration gap. Section~\ref{sec:framework} presents the design objectives, system functions, and two-layer architecture. Section~\ref{sec:components} details the five core technical components. Section~\ref{sec:validation} outlines the validation strategy. Section~\ref{sec:discussion} discusses trade-offs, limitations, and the path forward. Section~\ref{sec:conclusion} concludes the paper.

\begin{table*}[!t]
\def\arraystretch{1.2}
\centering
\caption{Commercial DHM platforms---integrated capability sets under proprietary licenses.}\label{tab:commercial}
\small
\begin{tabular}{@{}lx{2.4cm}x{4.0cm}x{4.0cm}x{2.0cm}@{}}
\hline \hline
\textbf{Tool} & \textbf{Developer} & \textbf{Key capabilities} & \textbf{Assessment methods} & \textbf{Focus} \\
\hline
Tecnomatix Jack~\cite{SiemensJack2017} & Siemens & 26-dim.\ anthro.\ scaling, task simulation, reach/vision, occupant packaging & Spinal compression, static strength, fatigue (base); NIOSH, RULA, OWAS, Snook (TAT add-on) & General industry \\
DELMIA~\cite{Polasek2015} & Dassault Syst\`{e}mes & Human Builder in CATIA/3DEXPERIENCE & NIOSH, Snook, biomechanical (L4/L5 compression, joint moments) & General industry \\
RAMSIS~\cite{VanDerMeulen2007} & Humanetics & Multi-country anthro.\ databases, probability-based posture prediction & Comfort assessment, packaging analysis & Vehicle interiors \\
Santos~\cite{AbdelMalek2007,Marler2009} & Univ.\ of Iowa & 209-DOF physics-based model, multi-objective optimization & Posture prediction, strength analysis & Military / defense \\
\hline \hline
\end{tabular}
\end{table*}

\begin{table*}[!b]
\def\arraystretch{1.2}
\centering
\caption{Open-source tools relevant to DHM---capable fragments without integration.}\label{tab:landscape}
\small
\begin{tabular}{@{}llx{4.2cm}x{4.2cm}@{}}
\hline \hline
\textbf{Category} & \textbf{Tool} & \textbf{Capabilities} & \textbf{What it lacks} \\
\hline
Musculoskeletal & OpenSim~\cite{Delp2007,Seth2018} & IK, ID, muscle forces, optimization-based motion prediction & No ergo.\ assessment, no CAD \\
 & biorbd~\cite{Michaud2021} & C++/Python musculoskeletal, IK/dynamics & Clinical focus, no workplace ergo. \\
 & AnyPyTools~\cite{Lund2019} & Python wrapper for AnyBody Modeling System & Underlying solver is commercial; no ergo.\ assessment, no CAD \\
Body generation & MakeHuman~\cite{Paul2019} & Parametric 3D mesh, morphological params, animation-grade skeleton rig & No anthropometric scaling, no biomechanical analysis \\
Physics engines & MuJoCo~\cite{Todorov2012} & Fast multi-body dynamics, humanoid models & No anthropometry, no ergo.\ scoring \\
Biomechanics & Pose2Sim~\cite{Pagnon2022} & Markerless $\rightarrow$ OpenSim pipeline & Capture tool, not design tool \\
 & pycapacity~\cite{Skuric2023} & Force/velocity polytopes & Narrow scope \\
Ergo.\ calculators & rs9000/ergonomics & Python REBA scoring & No mannequin, no CAD \\
CAD platform & FreeCAD & LGPL, OpenCASCADE, deep Python API & No DHM workbench \\
\hline \hline
\end{tabular}
\end{table*}

\section{Background and Gap Analysis}\label{sec:background}

\subsection{Commercial DHM Tools}\label{sec:commercial}

The commercial DHM market is defined by four platforms, each providing the integrated capability set but under proprietary licenses.

\textbf{Tecnomatix Jack} originated as a research system at the University of Pennsylvania~\cite{Badler1993} and was subsequently commercialized before becoming part of the Siemens Tecnomatix portfolio~\cite{Blanchonette2010}. Jack's anthropometric scaling system supports up to 26 anthropometric dimensions~\cite{Blanchonette2010} from built-in databases including ANSUR and NHANES~\cite{SiemensJack2017}. In addition to the base analyses documented in the Jack User Manual---low-back spinal compression analysis, 3D static strength prediction, and fatigue analysis~\cite{SiemensJack2017}---the separately licensed Task Analysis Toolkit (TAT) extends the suite with the NIOSH Lifting Equation, RULA, OWAS, Snook material handling limits, and metabolic energy expenditure analysis. The Task Simulation Builder enables sequencing of human operations with automatic posture recomputation; the Occupant Packaging Toolkit adds seated posture prediction and comfort assessment for vehicle design~\cite{SiemensOPT2017}. Validation studies documented anthropometric errors of 0.5--8.4\% and reach envelope mean differences of 1--2~cm~\cite{Blanchonette2010}.

\textbf{DELMIA} (Dassault Syst\`{e}mes) provides Human Builder within CATIA/3DEXPERIENCE; reported analysis modules include the NIOSH Lifting Equation, Snook/Ciriello manual handling limits, and biomechanical analysis (L4/L5 spinal compression and joint forces and moments)~\cite{Polasek2015}. \textbf{RAMSIS} (Humanetics) specializes in vehicle interior ergonomics with anthropometric databases from multiple countries and probability-based posture prediction~\cite{VanDerMeulen2007}. \textbf{Santos} (University of Iowa), developed with U.S.~Department of Defense funding~\cite{AbdelMalek2007}, provides a 209-DOF physics-based model with multi-objective posture prediction~\cite{Marler2009}.

Table~\ref{tab:commercial} summarizes the four platforms across key capability dimensions.

Beyond their capability sets, these tools share two practical characteristics relevant to this paper. Schall et al.~\cite{Schall2018} document licensing, training, and simulation development time as recurring cost factors in occupational-safety deployments. Their closed-source nature also means the underlying algorithms cannot be inspected or modified by end users, which is the condition that produces the cross-platform RWL differences reported by Pol\'{a}\v{s}ek et al.~\cite{Polasek2015} (Section~\ref{sec:introduction}).

\subsection{The Open-Source Landscape}\label{sec:opensource}

A systematic survey of open-source tools reveals a landscape of capable but disconnected components. Table~\ref{tab:landscape} summarizes the key tools across four capability categories.

The musculoskeletal tools (OpenSim, biorbd, AnyBody's open wrappers~\cite{Lund2019}) serve clinical and research biomechanics. MakeHuman produces visually plausible meshes with an animation-grade skeleton rig, but uses normalized morphological parameters rather than direct anthropometric dimensions and provides no biomechanical kinematics or ergonomic assessment. MuJoCo~\cite{Todorov2012} and PyBullet~\cite{CoumansBai2021} are general-purpose physics engines oriented toward robotics, model-based control, and machine learning rather than workplace ergonomic design. Among the published open-source biomechanics tools~\cite{Michaud2021,Pagnon2022,Skuric2023}, none addresses occupational ergonomic design evaluation within a CAD environment.

\subsection{The Integration Gap}\label{sec:gap}

The gap can be stated precisely: the open-source ecosystem has mature, individually validated components for each of the four core DHM capabilities, but \emph{no tool integrates them}. Table~\ref{tab:gap} frames the gap across the four capabilities.

Three stakeholder groups are directly affected by this gap:
\begin{list}{}{\setlength{\leftmargin}{0pt}\setlength{\itemindent}{0pt}\setlength{\listparindent}{0pt}\setlength{\itemsep}{2pt}\setlength{\parsep}{0pt}\setlength{\topsep}{4pt}}
\item \textbf{Ergonomics researchers} who need modifiable DHM for validation studies but currently pay commercial licenses or build single-use scripts (e.g., custom REBA calculators, one-off mannequin scaling routines, isolated IK solvers) that are neither reusable nor interoperable. Other domains have demonstrated that a shared open framework---ROS for robotics~\cite{Quigley2009}, PyTorch for deep learning~\cite{Paszke2019}---can unify fragmented efforts into a common platform; occupational ergonomics lacks such a framework.
\item \textbf{Small-to-medium enterprises} that cannot afford the per-seat licenses for tools such as Siemens Jack or Dassault DELMIA and consequently fall back on observational and mostly analog methods (e.g., manual RULA worksheets, pen-and-paper NIOSH calculations), forgoing the proactive, design-stage assessments that digital tools enable.
\item \textbf{Engineering educators} who cannot provide hands-on DHM experience without institutional licenses, limiting coursework to static images or video demonstrations rather than interactive posture analysis, DHM tutorials, and assessment exercises.
\end{list}

\section{The \openj{} Conceptual Framework}\label{sec:framework}

\subsection{Design Objectives}\label{sec:objectives}

Five design objectives trace from the identified gap and stakeholder needs:

\begin{list}{}{\setlength{\leftmargin}{0pt}\setlength{\itemindent}{0pt}\setlength{\listparindent}{0pt}\setlength{\itemsep}{2pt}\setlength{\parsep}{0pt}\setlength{\topsep}{4pt}}
\item \textbf{(DO-1) Integration.} Unify the four core DHM capabilities in a single open-source tool.
\item \textbf{(DO-2) Anthropometric fidelity.} Represent and scale a parametric mannequin from open population data within a CAD geometry framework.
\item \textbf{(DO-3) Posture prediction.} Provide plausible, interactive-speed posture prediction for workplace tasks.
\item \textbf{(DO-4) Assessment extensibility.} Support interchangeable ergonomic assessment methods with varying automation levels via a plugin architecture.
\item \textbf{(DO-5) Community contribution.} Fill the demonstrated open-source gap with well-documented, tested, and extensible software that serves the research and practitioner communities.
\end{list}

\begin{table}[!t]
\def\arraystretch{1.2}
\centering
\caption{The four-capability integration gap.}\label{tab:gap}
\small
\begin{tabular}{@{}lcc@{}}
\hline \hline
\textbf{Capability} & \textbf{Commercial} & \textbf{Open-source} \\
\hline
Scalable mannequin & \checkmark & Fragments \\
Posture prediction (IK) & \checkmark & Available \\
Ergonomic assessment & \checkmark & Scattered \\
CAD integration & \checkmark & None \\
\hline
\textbf{All four integrated} & \checkmark & --- \\
\hline \hline
\end{tabular}
\end{table}

\subsection{System Functions}\label{sec:functions}

Twelve functions decompose the design objectives (DO) into the operational capabilities required:

\begin{figure*}[!t]
\centering
\includegraphics[width=\textwidth]{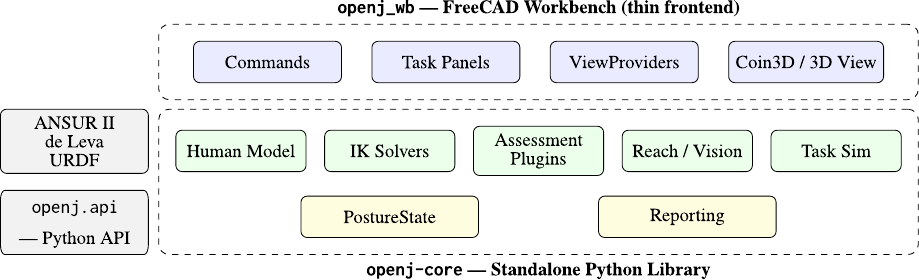}
\caption{Proposed two-layer architecture of \openj{}: a standalone core library (\texttt{openj-core}) and a thin FreeCAD workbench frontend (\texttt{openj\_wb}).}\label{fig:architecture}
\end{figure*}

\begin{list}{}{\setlength{\leftmargin}{0pt}\setlength{\itemindent}{0pt}\setlength{\listparindent}{0pt}\setlength{\itemsep}{2pt}\setlength{\parsep}{0pt}\setlength{\topsep}{4pt}}
\item \textbf{(F1)} Create human model (DO-2). Instantiate a parametric multi-segment mannequin with kinematic joints and body-segment parameters.
\item \textbf{(F2)} Scale to population percentile (DO-2). Adjust segment lengths and masses to match a target stature, weight, and sex using ANSUR~II data and de~Leva regression.
\item \textbf{(F3)} Predict posture via IK (DO-3). Solve for joint angles that place end-effectors at task-defined targets while minimizing a comfort-weighted objective.
\item \textbf{(F4)} Manipulate posture manually (DO-3). Allow interactive drag-based joint adjustment in the 3-D viewport with real-time differential IK.
\item \textbf{(F5)} Run ergonomic assessment (DO-4). Execute one or more pluggable assessment methods (e.g., RULA, REBA, NIOSH) on the current posture state.
\item \textbf{(F6)} Visualize risk on mannequin (DO-4). Map assessment scores to a color gradient rendered directly on body segments for immediate visual feedback.
\item \textbf{(F7)} Compute reach envelope (DO-1). Generate the reachable workspace volume via Monte Carlo sampling in joint space and convex-hull construction.
\item \textbf{(F8)} Compute vision cone (DO-1). Model the asymmetric human visual field and perform occlusion checks against scene geometry.
\item \textbf{(F9)} Import workplace geometry (DO-1). Load CAD assemblies (STEP, IGES, native FreeCAD) as the environment against which ergonomic analyses are performed.
\item \textbf{(F10)} Simulate task sequences (DO-1). Iterate through ordered action sequences (reach, grasp, move, place) and accumulate exposure scores over a shift.
\item \textbf{(F11)} Export assessment reports (DO-5). Generate structured output (JSON, CSV, PDF) summarizing posture data, scores, and risk levels for documentation.
\item \textbf{(F12)} Provide Python API (DO-5). Expose all core functionality as a scripting interface for batch analyses, CI pipelines, and third-party integration.
\end{list}

\subsection{Two-Layer Architecture}\label{sec:architecture}

Three architectural concepts were evaluated: (A)~a monolithic FreeCAD workbench, where all logic is embedded directly in the workbench---simple to bootstrap but untestable without a running FreeCAD instance and locked to a single frontend; (B)~a core library plus FreeCAD frontend, where computation resides in a standalone Python package and FreeCAD serves as a thin visualization/GUI layer; and (C)~a microservice architecture, where the core logic is exposed through a REST or gRPC API---maximizing language-agnostic integration but introducing network overhead, deployment complexity, and latency unsuitable for interactive posture manipulation. Concept~B was selected for its balance of testability, extensibility, and alignment with the core-library-plus-Python-bindings pattern used by biorbd~\cite{Michaud2021}. Concept~C is not mutually exclusive with Concept~B: a microservice layer can be added on top of the core library in future versions to serve web-based or multi-tool workflows without altering the two-layer foundation.

The architecture separates concerns into two packages:
\begin{list}{}{\setlength{\leftmargin}{0pt}\setlength{\itemindent}{0pt}\setlength{\listparindent}{0pt}\setlength{\itemsep}{2pt}\setlength{\parsep}{0pt}\setlength{\topsep}{4pt}}
\item \textbf{\texttt{openj-core}:} a standalone, pip-installable Python library with no FreeCAD dependency. Will contain the human model, IK solvers, assessment plugins, reach/vision computation, task simulation engine, and reporting. Testable via pytest and usable from scripts, Jupyter notebooks, or other frontends.
\item \textbf{\texttt{openj\_wb}:} a thin FreeCAD workbench that will call \texttt{openj-core} for all computation and handle 3D rendering, GUI panels, and user interaction through FreeCAD's Coin3D scenegraph and Qt-based panel system.
\end{list}

Figure~\ref{fig:architecture} illustrates the two-layer architecture. This separation will ensure that (1)~the core library can be tested independently without a GUI environment, (2)~researchers can use the Python API for scripted batch analyses, and (3)~future frontends (web-based, other CAD platforms) can be developed without modifying the core.

The FreeCAD workbench will follow established patterns from the Assembly, Draft, and RobotCAD workbenches~\cite{FreeCADWiki2024}: commands will be registered via \texttt{Gui.addCommand()}, task panels will use Qt \texttt{.ui} files, and 3D visualization will leverage Coin3D scenegraph nodes. A hybrid rendering strategy---direct Coin3D \texttt{SoTransform} manipulation during interactive posturing, with deferred document property updates---will be adopted from the approach demonstrated by Assembly4 and pivy\_trackers~\cite{Assembly4,PivyTrackers} to sustain 30+~FPS during drag manipulation.

\section{Core Technical Components}\label{sec:components}

\subsection{Anthropometric Human Model}\label{sec:human_model}

The \openj{} human model will be defined as a kinematic tree of 41 degrees of freedom: 6~DOF pelvis, 6~DOF spine (grouped lumbar and thoracic), 3~DOF neck, bilateral 3~DOF shoulders, 2~DOF elbows, 2~DOF wrists, 3~DOF hips, 1~DOF knees, and 2~DOF ankles. Finger articulation is excluded from the initial scope to bound complexity. The skeleton will be encoded in the Unified Robot Description Format (URDF)~\cite{Quigley2009}, which is natively supported by Pinocchio~\cite{Carpentier2019}, and augmented by a YAML sidecar file carrying DHM-specific metadata: comfort weights per joint, neutral posture angles, anthropometric scaling rules, and body segment parameter (BSP) source references.

Anthropometric scaling will follow a three-tier pipeline grounded in the ANSUR~II dataset~\cite{Gordon2014}:

\begin{list}{}{\setlength{\leftmargin}{0pt}\setlength{\itemindent}{0pt}\setlength{\listparindent}{0pt}\setlength{\itemsep}{2pt}\setlength{\parsep}{0pt}\setlength{\topsep}{4pt}}
\item \textbf{Tier~1 (direct measurement):} Where an ANSUR~II measurement maps directly to a segment length (e.g., upper arm length, thigh length), the measurement is used without transformation.
\item \textbf{Tier~2 (multivariate regression):} For segments without direct ANSUR~II equivalents, multivariate regression from stature, weight, and sitting height will predict segment dimensions, targeting $R^2 > 0.7$.
\item \textbf{Tier~3 (proportionality fallback):} For any remaining gaps, Winter/Drillis proportionality constants~\cite{Winter2009,Drillis1966} are applied as ratios of stature, with a logged warning indicating reduced accuracy.
\end{list}

Body segment parameters (segment mass, center of mass, and moment of inertia) will be computed from de~Leva~\cite{deLeva1996}, which provides sex-specific ratio-based parameters for 11~body segments as functions of total body mass and segment length, correcting earlier data from Zatsiorsky and Seluyanov~\cite{Zatsiorsky1983}.

Each body segment will be rendered as a geometric primitive (capsule or cylinder) scaled from segment dimensions. This approach, while visually simpler than the skinned vertex models used in computer vision (e.g., SMPL~\cite{Loper2015}), is sufficient for ergonomic evaluation and avoids the licensing complications of high-fidelity mesh assets.

\subsection{Posture Prediction}\label{sec:ik}

Posture prediction in DHM is formulated as a constrained optimization problem: given task constraints (reach targets, external loads, support conditions), find joint angles that minimize a comfort objective while respecting joint limits, balance, and collision constraints. This optimization-based paradigm, established by Jack~\cite{Zhao1994} and extensively developed in the Santos framework~\cite{Marler2009,Marler2005,Xiang2010}, remains one of the two principal approaches in DHM---alongside data-driven methods---as documented by a recent systematic review of 24~studies that found an even split between the two paradigms~\cite{Zhang2025}.

\openj{} will provide two complementary solvers:

\textbf{Optimization-based posture prediction.} A comfort-weighted objective function
\begin{equation}\label{eq:comfort}
    \min_{\mathbf{q}} \sum_{j=1}^{n} w_j \left(q_j - q_j^{\text{neutral}}\right)^2
\end{equation}
subject to end-effector position constraints, joint limit bounds $q_j^{\min} \leq q_j \leq q_j^{\max}$, and optional balance constraints (center-of-mass projection within the support polygon). Default joint comfort weights $w_j$ and neutral angles $q_j^{\text{neutral}}$ will be drawn from population-average values reported in the Santos discomfort literature~\cite{Marler2005,Marler2009}; users will be able to override per-individual values via the YAML metadata sidecar. The solver will use Sequential Least Squares Programming (SLSQP) via SciPy~\cite{Virtanen2020}, matching the general formulation of Santos~\cite{Marler2009} while remaining accessible as a pure-Python implementation.

\textbf{Differential IK for interactive manipulation.} For real-time posture adjustment during drag-based interaction in the FreeCAD viewport, \openj{} will employ Pinocchio~\cite{Carpentier2019} with the Pink framework~\cite{Caron2022}, which provides QP-based differential IK with weighted task objectives built on top of Pinocchio. A pure-Python fallback using SciPy will be provided for environments where Pinocchio's C++ dependencies cannot be installed.

\subsection{Pluggable Ergonomic Assessment Framework}\label{sec:assessment}

Ergonomic assessment methods vary in their automation potential. The semi-automatic platform of Generosi et al.~\cite{Generosi2022} extracts posture angles from skeletal data and partially fills the RULA, REBA, OCRA, and OWAS checklists automatically, while still requiring the ergonomist to enter the remaining checklist items. Methods such as the NIOSH Lifting Equation and Snook tables go further: their inputs (load weight, horizontal/vertical reach distances, frequency, coupling quality) cannot be inferred from joint angles at all and must be user-supplied. \openj{} will accommodate this spectrum through a plugin architecture.

Each assessment method will implement an abstract base class:

\begin{verbatim}
class ErgonomicAssessment(ABC):
    @abstractmethod
    def assess(self, state: PostureState)
        -> AssessmentResult: ...

    @property
    def automation_level(self)
        -> str: ...  # FULL | PARTIAL

    @property
    def required_context_fields(self)
        -> list[str]: ...
\end{verbatim}

The \texttt{PostureState} dataclass will provide the canonical input: joint angles (radians, URDF convention), ergonomic angles (degrees, anatomical convention---trunk flexion, upper arm abduction, etc.), support type (standing, sitting), and a context dictionary for method-specific parameters. The framework will validate required context fields before execution and raise a descriptive error if any are missing.

Five built-in plugins will implement the most widely used industrial assessment methods:

\begin{list}{}{\setlength{\leftmargin}{0pt}\setlength{\itemindent}{0pt}\setlength{\listparindent}{0pt}\setlength{\itemsep}{2pt}\setlength{\parsep}{0pt}\setlength{\topsep}{4pt}}
\item \textbf{RULA}~\cite{McAtamney1993}: Scores 1--7 across four action levels for upper-limb-intensive tasks.
\item \textbf{REBA}~\cite{Hignett2000}: Scores 1--15 across five risk levels for whole-body assessment.
\item \textbf{OWAS}~\cite{Karhu1977}: Classifies postures into four action categories using back, upper-limb, and lower-limb codes.
\item \textbf{NIOSH Lifting Equation}~\cite{Waters1993}: Computes the Recommended Weight Limit (RWL) and Lifting Index (LI) from six task variables.
\item \textbf{Snook/Liberty Mutual tables}~\cite{Snook1991}: Compares observed task demands against population-specific acceptable limits for lift, lower, push, pull, and carry actions.
\end{list}

Because the plugin interface will be defined as a Python abstract base class, researchers will be able to implement custom assessment methods---novel scoring systems, domain-specific indices, or experimental metrics---without modifying the framework core. This extensibility directly addresses the needs of ergonomics researchers who require the ability to test new assessment approaches within an integrated DHM environment.

\begin{table*}[t]
\def\arraystretch{1.2}
\centering
\caption{Validation plan for core \openj{} components.}\label{tab:validation}
\small
\begin{tabular}{@{}x{3.2cm}x{2.5cm}x{6cm}x{2.8cm}@{}}
\hline \hline
\textbf{Component} & \textbf{Method} & \textbf{Acceptance criterion} & \textbf{Reference} \\
\hline
Anthropometric scaling & Numerical test & Direct ANSUR~II measurements within 1\,mm; regression predictions $R^2 > 0.7$; BSP values within 1\% relative error of de~Leva & \cite{Gordon2014,deLeva1996} \\
Posture prediction (IK) & Numerical test & End-effector within 5\,mm of target; all joint angles within limits; differential IK mean $<$\,33\,ms (30+\,FPS) & \cite{Marler2009,Carpentier2019,Caron2022} \\
RULA / REBA / OWAS & Comparison test & $\geq$5 worked postures per method drawn from primary publications; exact grand-score match required for $\geq$80\% of cases, $\pm$1 point allowed otherwise & \cite{McAtamney1993,Hignett2000,Karhu1977} \\
NIOSH / Snook & Comparison test & $\geq$5 scenarios with documented inputs from primary publications; RWL/LI within $\pm$2\% of the equation evaluated by hand on the same inputs & \cite{Waters1993,Snook1991} \\
Reach envelope & Comparison test & End-effector reach surfaces within mean differences of 1--2\,cm relative to empirical pilot reach data reported for Jack & \cite{Blanchonette2010} \\
\hline \hline
\end{tabular}
\end{table*}

\subsection{Reach and Vision Analysis}\label{sec:reach_vision}

Reach envelope computation will use Monte Carlo sampling in joint space: joint angles are drawn uniformly within limits, forward kinematics maps each configuration to an end-effector position, and a convex hull~\cite{Virtanen2020} of the resulting point cloud defines the reachable workspace volume. This approach is computationally simple and parallelizable; for validation, predicted reach surfaces will be compared against the empirical mean reach differences of approximately 1--2~cm reported for Jack against pilot reach data~\cite{Blanchonette2010}.

Vision analysis will adopt an asymmetric cone model informed by Vinayak and Sen~\cite{VinayakSen2012}, who demonstrated that the human visual field is wider temporally (toward the ear) than nasally (toward the nose). The central cone will be set to approximately 60\textdegree{} (functional visual field) by default, and the peripheral cone to approximately 120\textdegree{} (awareness-level perception); both are user-configurable per analysis to accommodate task-specific or individual values. Occlusion checking will use ray-casting against scene geometry provided by FreeCAD's OpenCASCADE kernel.

\subsection{Task Simulation and Cumulative Exposure}\label{sec:task_sim}

Single-posture ergonomic assessments, while valuable, do not capture the cumulative effect of repeated exposures over a work shift. \openj{} will address this through a lightweight task simulation engine. Tasks will be defined in JSON or YAML as ordered sequences of actions---reach, grasp, move, place, hold---each with target positions, loads, and durations. The engine will iterate through the sequence, solving IK for each step, running selected assessments, and computing cumulative scores including time-weighted REBA exposure and the NIOSH Composite Lifting Index for multi-task lifting~\cite{Waters1994}. This approach follows the paradigm of Jack's Task Simulation Builder~\cite{SiemensJack2017} while using an open, human-readable task definition format.

Emerging work on cumulative damage modeling~\cite{Gallagher2017} suggests that fatigue-failure analogies (S-N curves applied to biomechanical loading) may eventually supplement categorical risk scores. The \openj{} architecture will accommodate such extensions through the plugin framework.

\section{Validation Strategy}\label{sec:validation}

Validation of an ergonomic assessment framework requires demonstrating that (1)~the human model is anthropometrically accurate, (2)~posture prediction produces biomechanically plausible results, and (3)~assessment scores match published reference values. Table~\ref{tab:validation} summarizes the planned validation approach across the five core components. The acceptance criteria are stated as initial design targets to guide implementation; final tolerances will be refined once the corresponding components are built and benchmark results accumulate.


All validation tests will be automated via pytest and enforced through continuous integration (GitHub Actions) across three platforms (Ubuntu, Windows, macOS). A minimum of 80\% code coverage is targeted for the core library, following open-source software quality best practices.

\begin{figure*}[!b]
\centering
\includegraphics[width=\textwidth]{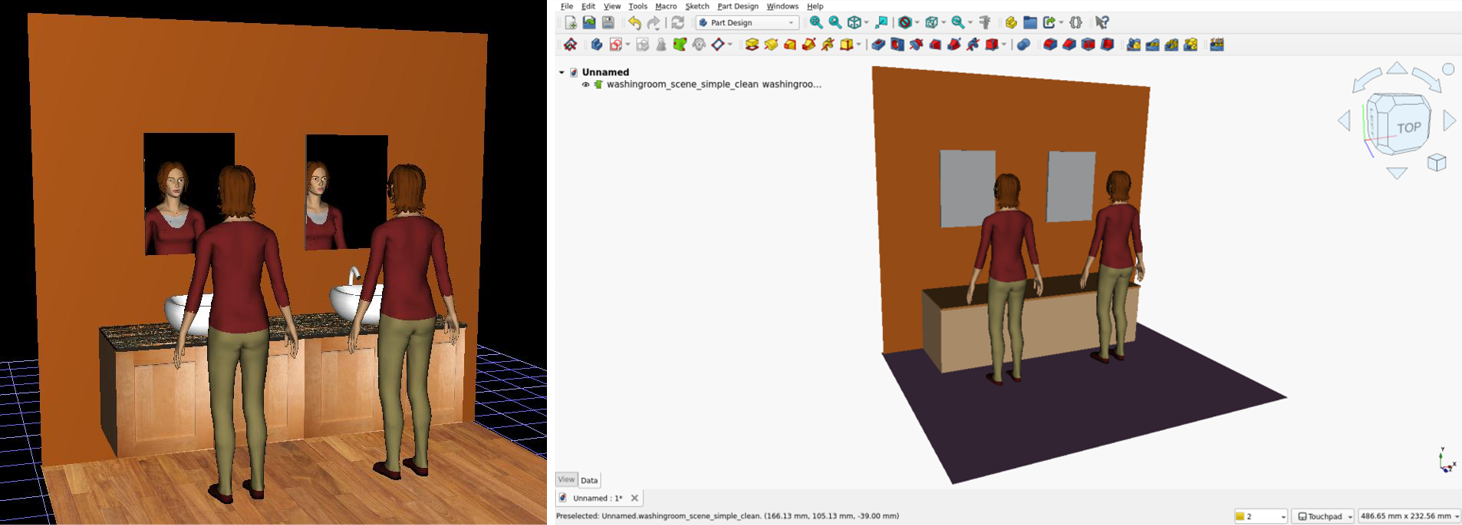}
\caption{Cross-tool scene: a tutorial scene from Tecnomatix Jack Student Tutorial~\cite{SiemensJack2013} (left) and its simplified representation in FreeCAD (right).}
\label{fig:jack_freecad}
\end{figure*}

As a preliminary feasibility check, we utilized an existing tutorial scene from Tecnomatix Jack Student Tutorial and imported it in FreeCAD (Fig.~\ref{fig:jack_freecad}). The visualization shows that the environment geometry and human figure positions transfer successfully while excluding some of the existing software-dependent features of the Jack (e.g., facet reflection) and this supports the potential use of FreeCAD as a host platform for \openj{}.

\section{Discussion}\label{sec:discussion}

\subsection{Design Trade-offs}\label{sec:tradeoffs}

Several design decisions reflect deliberate trade-offs between capability and tractability for an open-source project:

\textbf{Geometric primitives vs.\ skinned meshes.} The decision to render body segments as capsules and cylinders rather than SMPL-derived skinned meshes~\cite{Loper2015} sacrifices visual realism for licensing clarity, reduced dependency complexity, and computational simplicity. Ergonomic assessments depend on joint angles and segment dimensions, not surface detail; the primitive approach is sufficient for all planned assessment methods. Future versions could use CC0-licensed MakeHuman exports~\cite{Paul2019} to improve visual realism for educational or presentation contexts, while retaining the parametric primitive geometry as the analysis target.

\textbf{41 DOF vs.\ full articulation.} Commercial tools such as Jack model 69~joints with 135~DOF across 71~segments, with a fully articulated hand and spine model~\cite{SiemensJack2017}. \openj{} will group the spine into 6~DOF and omit fingers in the initial scope, substantially reducing URDF authoring effort and IK solve time while retaining the degrees of freedom relevant to industrial ergonomic assessment methods (which score trunk flexion, not individual vertebral angles).

\textbf{Optimization-based IK vs.\ data-driven prediction.} While neural-network surrogates can predict postures in fractions of a second~\cite{Bataineh2016,Zhang2025}, they require large training datasets that do not yet exist for the general DHM use case. The optimization-based approach will provide biomechanical transparency---every predicted posture can be explained in terms of the objective function and constraints---at the cost of somewhat longer solve times for the offline solver. The differential IK path (Pinocchio + Pink) will provide interactive speed for the FreeCAD viewport.

\subsection{Positioning Relative to Existing Tools}\label{sec:positioning}

\openj{} is not intended to compete with musculoskeletal simulators; it aims to complement them. OpenSim~\cite{Delp2007,Seth2018} and biorbd~\cite{Michaud2021} excel at inverse dynamics, muscle force estimation, and optimization-based motion prediction---capabilities that \openj{} will not attempt to replicate. Instead, \openj{} is designed to fill the \emph{occupational ergonomic design evaluation} niche: placing a mannequin in a CAD workplace, predicting task postures, and scoring them against industrial standards. The two tool categories serve different stages of the design-to-analysis pipeline, and future interoperability (e.g., exporting \openj{} postures to OpenSim for detailed biomechanical analysis) would be an intuitive extension.

Beyond its core role as a design tool, \openj{} can serve two additional research functions. First, its transparent, open-source implementation enables independent validation studies: researchers can inspect, reproduce, and compare ergonomic scoring algorithms against published reference values or commercial tool outputs, providing a means to investigate the cross-platform RWL differences reported by Pol\'{a}\v{s}ek et al.~\cite{Polasek2015}. Second, the integration of posture prediction with CAD geometry enables proactive, design-stage simulation of work tasks before physical implementation, allowing engineers to identify and mitigate ergonomic risks during the design phase rather than through post-hoc observational assessment of existing workplaces.

\openj{} addresses a different purpose from vision-based and sensor-based ergonomic assessment pipelines such as those of Generosi et al.~\cite{Generosi2022} and Gonz\'{a}lez-Alonso et al.~\cite{GonzalezAlonso2024}. Vision/IMU approaches operate on workers physically performing the task and serve observational, post-deployment, and continuous-monitoring use cases. \openj{} targets design-stage assessment---evaluating a workstation in CAD before workers are present or production lines are built. The two purposes are not mutually exclusive and can be mixed within a single use case: for example, \openj{} can be used for design-stage evaluation while a vision/IMU pipeline performs post-deployment validation, or observed postures from a vision pipeline can be fed into \openj{} to score what-if redesigns of an existing workstation.

\subsection{Limitations and Future Work}\label{sec:limitations}

The framework as presented has several acknowledged limitations:

\begin{list}{}{\setlength{\leftmargin}{0pt}\setlength{\itemindent}{0pt}\setlength{\listparindent}{0pt}\setlength{\itemsep}{2pt}\setlength{\parsep}{0pt}\setlength{\topsep}{4pt}}
\item \textbf{Population representation.} ANSUR~II represents a U.S.~military population, which is not representative of civilian populations globally. Integration of additional anthropometric databases (NHANES, WHO growth references, national surveys) is planned for future versions.
\item \textbf{Reduced articulation.} Because the framework omits finger articulation and groups the spine into 6~DOF, certain assessment inputs cannot be inferred from posture alone---specifically, the wrist/hand subscores in RULA, the coupling-quality input in REBA, and the coupling multiplier (CM) in NIOSH must be user-supplied. The plugin framework's \texttt{required\_context\_fields} mechanism handles these declaratively.
\item \textbf{Collision detection.} Self-collision and mannequin-environment collision during IK are not addressed in the initial scope. FreeCAD's OpenCASCADE kernel provides \texttt{distToShape()} for clearance analysis, which could be integrated as a post-solve check.
\item \textbf{Subjective factors.} Like all DHM tools, \openj{} will not capture subjective comfort, thermal environment, psychosocial stressors, or cognitive load. These factors fall outside the biomechanical scope of the framework.
\item \textbf{No real-time motion capture integration.} The framework will operate on defined postures and tasks; integration with markerless motion capture (e.g., Pose2Sim~\cite{Pagnon2022}) for real-time ergonomic monitoring is deferred to future work.
\end{list}

The immediate next step is implementation following a vertical-slice strategy: the FreeCAD workbench is scaffolded first, and each subsequent phase delivers core logic and its GUI simultaneously---producing a visible, testable mannequin from the first development sprint. The completed tool will be released as an open-source package with comprehensive documentation, automated testing, and community contribution guidelines to encourage adoption and collaborative development.

\section{Conclusion}\label{sec:conclusion}

This paper presents a conceptual framework for \openj{}, an open-source digital human modeling and ergonomic assessment tool intended to bridge a well-documented gap in the open-source ecosystem. Our literature survey confirms that no existing open-source tool integrates the four core capabilities required for computational ergonomic workplace design: a scalable anthropometric mannequin, posture prediction, pluggable ergonomic assessments, and CAD environment integration.

The proposed \openj{} framework addresses this gap through a two-layer architecture---a standalone Python core library and a thin FreeCAD workbench frontend---targeting three underserved communities: researchers who need modifiable and reproducible analyses, industrial engineers who need zero-cost ergonomic evaluation, and educators who need hands-on DHM instruction. All major design decisions are traced to peer-reviewed literature, and a validation plan with quantitative acceptance criteria is specified to guide future implementation.

By publishing this conceptual framework prior to implementation, we seek community input on priorities, architecture, and validation approaches. The \openj{} project will follow an open development model, and contributions from the ergonomics, biomechanics, and FreeCAD communities will be welcomed.


\bibliographystyle{asmeconf}
\bibliography{References}

@book{Badler1993,
  author    = {Badler, Norman I. and Phillips, Cary B. and Webber, Bonnie L.},
  title     = {\href{https://doi.org/10.1093/oso/9780195073591.001.0001}{Simulating Humans: Computer Graphics Animation and Control}},
  publisher = {Oxford University Press},
  address   = {New York},
  year      = {1993},
}

@techreport{Blanchonette2010,
  author      = {Blanchonette, Peter},
  title       = {\href{https://apps.dtic.mil/sti/html/tr/ADA518132/}{Jack Human Modelling Tool: A Review}},
  institution = {Defence Science and Technology Organisation (DSTO)},
  number      = {DSTO-TR-2364},
  year        = {2010},
  address     = {Victoria, Australia},
}

@manual{SiemensJack2013,
  author = {{Siemens Product Lifecycle Management Software Inc.}},
  title  = {Jack 8.0.1 Student Edition Tutorial},
  year   = {2013},
}

@manual{SiemensJack2017,
  author = {{Siemens Product Lifecycle Management Software Inc.}},
  title  = {Jack User Manual, Version~9.0},
  year   = {2017},
  note   = {Document MT60010-S-084},
}

@manual{SiemensOPT2017,
  author = {{Siemens Product Lifecycle Management Software Inc.}},
  title  = {Jack Occupant Packaging Toolkit ({OPT}) Training Manual, Version~9.0},
  year   = {2017},
}

@book{Chaffin2001,
  author    = {Chaffin, Don B.},
  title     = {\href{https://colostate.primo.exlibrisgroup.com/permalink/01COLSU_INST/5c577e/cdi_proquest_miscellaneous_200152244}{Digital Human Modeling for Vehicle and Workplace Design}},
  publisher = {SAE International},
  year      = {2001},
  doi       = {10.4271/R-274},
}

@article{Chaffin2005,
  author  = {Chaffin, Don B.},
  title   = {\href{https://doi.org/10.1080/00140130400029191}{Improving Digital Human Modelling for Proactive Ergonomics in Design}},
  journal = {Ergonomics},
  volume  = {48},
  number  = {5},
  pages   = {478--491},
  year    = {2005},
  doi     = {10.1080/00140130400029191},
}

@article{AbdelMalek2007,
  author  = {Abdel-Malek, Karim and Yang, Jingzhou and Marler, Timothy and Beck, Steven and Mathai, Anith and Zhou, Xianlian and Patrick, Amos and Arora, Jasbir},
  title   = {\href{https://doi.org/10.1504/IJHFMS.2006.011680}{Towards a New Generation of Virtual Humans}},
  journal = {International Journal of Human Factors Modelling and Simulation},
  volume  = {1},
  number  = {1},
  pages   = {2--39},
  year    = {2006},
  doi     = {10.1504/IJHFMS.2006.011680},
}

@article{VanDerMeulen2007,
  author  = {van der Meulen, Peter and Seidl, Andreas},
  title   = {\href{https://doi.org/10.1007/978-3-540-73321-8_113}{{RAMSIS}---The Leading {CAD} Tool for Ergonomic Analysis of Vehicles}},
  journal = {Lecture Notes in Computer Science},
  volume  = {4561},
  pages   = {1008--1017},
  year    = {2007},
  doi     = {10.1007/978-3-540-73321-8_113},
}

@article{Polasek2015,
  author  = {Pol\'{a}\v{s}ek, Patrik and Bure\v{s}, Marek and \v{S}imon, Michal},
  title   = {\href{https://doi.org/10.1016/j.proeng.2015.01.494}{Comparison of Digital Tools for Ergonomics in Practice}},
  journal = {Procedia Engineering},
  volume  = {100},
  pages   = {1277--1285},
  year    = {2015},
  doi     = {10.1016/j.proeng.2015.01.494},
}

@article{Schall2018,
  author  = {Schall Jr, Mark C. and Fethke, Nathan B. and Roemig, Victoria},
  title   = {\href{https://doi.org/10.1080/24725838.2018.1491430}{Digital Human Modeling in the Occupational Safety and Health Process: An Application in Manufacturing}},
  journal = {IISE Transactions on Occupational Ergonomics and Human Factors},
  volume  = {6},
  number  = {2},
  pages   = {64--75},
  year    = {2018},
  doi     = {10.1080/24725838.2018.1491430},
}

@article{GonzalezAlonso2024,
  author  = {Gonz\'{a}lez-Alonso, Javier and Sim\'{o}n-Mart\'{i}nez, Cristina and Ant\'{o}n-Rodr\'{i}guez, Mar\'{i}a and Gonz\'{a}lez-Ortega, David and D\'{i}az-Pernas, Francisco J. and Mart\'{i}nez-Zarzuela, Mario},
  title   = {\href{https://doi.org/10.1016/j.ssci.2024.106431}{Development of an End-to-End Hardware and Software Pipeline for Affordable and Feasible Ergonomics Assessment in the Automotive Industry}},
  journal = {Safety Science},
  volume  = {173},
  pages   = {106431},
  year    = {2024},
  doi     = {10.1016/j.ssci.2024.106431},
}

@article{Delp2007,
  author  = {Delp, Scott L. and Anderson, Frank C. and Arnold, Allison S. and Loan, Peter and Habib, Ayman and John, Chand T. and Guendelman, Eran and Thelen, Darryl G.},
  title   = {\href{https://doi.org/10.1109/TBME.2007.901024}{{OpenSim}: Open-Source Software to Create and Analyze Dynamic Simulations of Movement}},
  journal = {IEEE Transactions on Biomedical Engineering},
  volume  = {54},
  number  = {11},
  pages   = {1940--1950},
  year    = {2007},
  doi     = {10.1109/TBME.2007.901024},
}

@article{Seth2018,
  author  = {Seth, Ajay and Hicks, Jennifer L. and Uchida, Thomas K. and Habib, Ayman and Dembia, Christopher L. and Dunne, James J. and Ong, Carmichael F. and DeMers, Matthew S. and Rajagopal, Apoorva and Millard, Matthew and others},
  title   = {\href{https://doi.org/10.1371/journal.pcbi.1006223}{{OpenSim}: Simulating Musculoskeletal Dynamics and Neuromuscular Control to Study Human and Animal Movement}},
  journal = {PLoS Computational Biology},
  volume  = {14},
  number  = {7},
  pages   = {e1006223},
  year    = {2018},
  doi     = {10.1371/journal.pcbi.1006223},
}

@article{Michaud2021,
  author  = {Michaud, Benjamin and Begon, Micka\"{e}l},
  title   = {\href{https://doi.org/10.21105/joss.02562}{{biorbd}: A {C++}, {Python} and {MATLAB} Library to Analyze and Simulate the Human Body Biomechanics}},
  journal = {Journal of Open Source Software},
  volume  = {6},
  number  = {57},
  pages   = {2562},
  year    = {2021},
  doi     = {10.21105/joss.02562},
}

@article{Pagnon2022,
  author  = {Pagnon, David and Domalain, Mathieu and Reveret, Lionel},
  title   = {\href{https://doi.org/10.3390/s21196530}{{Pose2Sim}: An End-to-End Workflow for {3D} Markerless Sports Kinematics---Part 1: Robustness}},
  journal = {Sensors},
  volume  = {21},
  number  = {19},
  pages   = {6530},
  year    = {2021},
  doi     = {10.3390/s21196530},
}

@article{Skuric2023,
  author  = {Skuric, Antun and Padois, Vincent and Daney, David},
  title   = {\href{https://doi.org/10.21105/joss.05670}{{pycapacity}: A Real-Time Task-Space Capacity Calculation Package for Robotics and Biomechanics}},
  journal = {Journal of Open Source Software},
  volume  = {8},
  number  = {89},
  pages   = {5670},
  year    = {2023},
  doi     = {10.21105/joss.05670},
}

@article{Lund2019,
  author  = {Lund, Morten Enemark and Rasmussen, John and Andersen, Michael Skipper},
  title   = {\href{https://doi.org/10.21105/joss.01108}{{AnyPyTools}: A {Python} Package for Reproducible Research with the {AnyBody Modeling System}}},
  journal = {Journal of Open Source Software},
  volume  = {4},
  number  = {33},
  pages   = {1108},
  year    = {2019},
  doi     = {10.21105/joss.01108},
}

@inproceedings{Todorov2012,
  author    = {Todorov, Emanuel and Erez, Tom and Tassa, Yuval},
  title     = {\href{https://doi.org/10.1109/IROS.2012.6386109}{{MuJoCo}: A Physics Engine for Model-Based Control}},
  booktitle = {IEEE/RSJ International Conference on Intelligent Robots and Systems},
  pages     = {5026--5033},
  year      = {2012},
  doi       = {10.1109/IROS.2012.6386109},
}

@misc{CoumansBai2021,
  author = {Coumans, Erwin and Bai, Yunfei},
  title  = {\href{https://pybullet.org}{{PyBullet}, a Python Module for Physics Simulation for Games, Robotics and Machine Learning}},
  year   = {2016--2021},
  url    = {https://pybullet.org},
}

@book{Paul2019,
  author    = {Scataglini, Sofia and Paul, Gunther},
  title     = {\href{https://colostate.primo.exlibrisgroup.com/permalink/01COLSU_INST/5j5t2p/alma991031725636603361}{{DHM} and Posturography}},
  publisher = {Academic Press},
  year      = {2019},
  doi       = {10.1016/C2018-0-00519-1},
}

@article{Loper2015,
  author  = {Loper, Matthew and Mahmood, Naureen and Romero, Javier and Pons-Moll, Gerard and Black, Michael J.},
  title   = {\href{https://doi.org/10.1145/2816795.2818013}{{SMPL}: A Skinned Multi-Person Linear Model}},
  journal = {ACM Transactions on Graphics},
  volume  = {34},
  number  = {6},
  pages   = {248:1--248:16},
  year    = {2015},
  doi     = {10.1145/2816795.2818013},
}

@techreport{Gordon2014,
  author      = {Gordon, Claire C. and Blackwell, Cynthia L. and Bradtmiller, Bruce and Parham, Joseph L. and Barrientos, Patricia and Paquette, Steven P. and Corner, Brian D. and Carson, Jeremy M. and Venezia, Joseph C. and Rockwell, Belva M. and Mucher, Michael and Kristensen, Stacy},
  title       = {\href{https://apps.dtic.mil/sti/html/tr/ADA611869/}{2012 Anthropometric Survey of {U.S.} Army Personnel: Methods and Summary Statistics}},
  institution = {U.S. Army Natick Soldier Research, Development and Engineering Center},
  number      = {TR-15-007},
  year        = {2014},
}

@article{deLeva1996,
  author  = {de Leva, Paolo},
  title   = {\href{https://doi.org/10.1016/0021-9290(95)00178-6}{Adjustments to {Zatsiorsky--Seluyanov's} Segment Inertia Parameters}},
  journal = {Journal of Biomechanics},
  volume  = {29},
  number  = {9},
  pages   = {1223--1230},
  year    = {1996},
  doi     = {10.1016/0021-9290(95)00178-6},
}

@book{Winter2009,
  author    = {Winter, David A.},
  title     = {\href{https://colostate.primo.exlibrisgroup.com/permalink/01COLSU_INST/5j5t2p/alma991020244019703361}{Biomechanics and Motor Control of Human Movement}},
  edition   = {4th},
  publisher = {John Wiley \& Sons},
  year      = {2009},
}

@techreport{Drillis1966,
  author      = {Drillis, Rudolfs and Contini, Renato},
  title       = {\href{https://yumpu.com/en/document/view/37732232/drills-contini-1966-body-segment-parameters}{Body Segment Parameters}},
  institution = {School of Engineering and Science, New York University},
  number      = {1166-03},
  year        = {1966},
}

@incollection{Zatsiorsky1983,
  author    = {Zatsiorsky, Vladimir and Seluyanov, V.},
  title     = {\href{https://www.researchgate.net/publication/263065659}{The Mass and Inertia Characteristics of the Main Segments of the Human Body}},
  booktitle = {Biomechanics VIII-B},
  publisher = {Human Kinetics},
  address   = {Champaign, IL},
  pages     = {1152--1159},
  year      = {1983},
}

@article{Zhao1994,
  author  = {Zhao, Jianmin and Badler, Norman I.},
  title   = {\href{https://doi.org/10.1145/195826.195827}{Inverse Kinematics Positioning Using Nonlinear Programming for Highly Articulated Figures}},
  journal = {ACM Transactions on Graphics},
  volume  = {13},
  number  = {4},
  pages   = {313--336},
  year    = {1994},
  doi     = {10.1145/195826.195827},
}

@article{Marler2009,
  author  = {Marler, R. Timothy and Arora, Jasbir S. and Yang, Jingzhou and Kim, Hyung-Joo and Abdel-Malek, Karim},
  title   = {\href{https://doi.org/10.1080/03052150902853013}{Use of Multi-Objective Optimization for Digital Human Posture Prediction}},
  journal = {Engineering Optimization},
  volume  = {41},
  number  = {10},
  pages   = {925--943},
  year    = {2009},
  doi     = {10.1080/03052150902853013},
}

@article{Marler2005,
  author  = {Marler, R. Timothy and Rahmatalla, Salam and Shanahan, Meagan and Abdel-Malek, Karim},
  title   = {\href{https://doi.org/10.4271/2005-01-2680}{A New Discomfort Function for Optimization-Based Posture Prediction}},
  journal = {SAE Technical Paper},
  number  = {2005-01-2680},
  year    = {2005},
  doi     = {10.4271/2005-01-2680},
}

@article{Xiang2010,
  author  = {Xiang, Yujiang and Arora, Jasbir S. and Abdel-Malek, Karim},
  title   = {\href{https://doi.org/10.1007/s00158-010-0496-8}{Physics-Based Modeling and Simulation of Human Walking: A Review of Optimization-Based and Other Approaches}},
  journal = {Structural and Multidisciplinary Optimization},
  volume  = {42},
  number  = {1},
  pages   = {1--23},
  year    = {2010},
  doi     = {10.1007/s00158-010-0496-8},
}

@article{Bataineh2016,
  author  = {Bataineh, Mohammad and Marler, Timothy and Abdel-Malek, Karim and Arora, Jasbir},
  title   = {\href{https://doi.org/10.1016/j.eswa.2015.11.020}{Neural Network for Dynamic Human Motion Prediction}},
  journal = {Expert Systems with Applications},
  volume  = {48},
  pages   = {26--34},
  year    = {2016},
  doi     = {10.1016/j.eswa.2015.11.020},
}

@article{Zhang2025,
  author  = {Zhang, Mengjie and Nieuwenhuys, Arne and Zhang, Yanxin},
  title   = {\href{https://doi.org/10.1016/j.medengphy.2025.104391}{Posture Prediction Models in Digital Human Modeling for Ergonomic Design: A Systematic Review}},
  journal = {Medical Engineering \& Physics},
  volume  = {143},
  number  = {1},
  pages   = {104391},
  year    = {2025},
  doi     = {10.1016/j.medengphy.2025.104391},
}

@inproceedings{Carpentier2019,
  author    = {Carpentier, Justin and Saurel, Guilhem and Buondonno, Gabriele and Mirabel, Joseph and Lamiraux, Florent and Stasse, Olivier and Mansard, Nicolas},
  title     = {\href{https://laas.hal.science/hal-01866228v2}{The {Pinocchio} {C++} Library: A Fast and Flexible Implementation of Rigid Body Dynamics Algorithms and Their Analytical Derivatives}},
  booktitle = {2019 IEEE/SICE International Symposium on System Integration (SII)},
  pages     = {614--619},
  year      = {2019},
  doi       = {10.1109/SII.2019.8700380},
}

@misc{Caron2022,
  author = {Caron, St\'{e}phane},
  title  = {\href{https://github.com/stephane-caron/pink}{Pink: Python Inverse Kinematics Based on {Pinocchio}}},
  year   = {2022},
  url    = {https://github.com/stephane-caron/pink},
}

@article{Virtanen2020,
  author  = {Virtanen, Pauli and Gommers, Ralf and Oliphant, Travis E. and Haberland, Matt and Reddy, Tyler and Cournapeau, David and Burovski, Evgeni and Peterson, Pearu and Weckesser, Warren and Bright, Jonathan and others},
  title   = {\href{https://doi.org/10.1038/s41592-019-0686-2}{{SciPy} 1.0: Fundamental Algorithms for Scientific Computing in {Python}}},
  journal = {Nature Methods},
  volume  = {17},
  number  = {3},
  pages   = {261--272},
  year    = {2020},
  doi     = {10.1038/s41592-019-0686-2},
}

@article{McAtamney1993,
  author  = {McAtamney, Lynn and Corlett, E. Nigel},
  title   = {\href{https://doi.org/10.1016/0003-6870(93)90080-S}{{RULA}: A Survey Method for the Investigation of Work-Related Upper Limb Disorders}},
  journal = {Applied Ergonomics},
  volume  = {24},
  number  = {2},
  pages   = {91--99},
  year    = {1993},
  doi     = {10.1016/0003-6870(93)90080-S},
}

@article{Hignett2000,
  author  = {Hignett, Sue and McAtamney, Lynn},
  title   = {\href{https://doi.org/10.1016/S0003-6870(99)00039-3}{Rapid Entire Body Assessment ({REBA})}},
  journal = {Applied Ergonomics},
  volume  = {31},
  number  = {2},
  pages   = {201--205},
  year    = {2000},
  doi     = {10.1016/S0003-6870(99)00039-3},
}

@article{Waters1993,
  author  = {Waters, Thomas R. and Putz-Anderson, Vern and Garg, Arun and Fine, Lawrence J.},
  title   = {\href{https://doi.org/10.1080/00140139308967940}{Revised {NIOSH} Equation for the Design and Evaluation of Manual Lifting Tasks}},
  journal = {Ergonomics},
  volume  = {36},
  number  = {7},
  pages   = {749--776},
  year    = {1993},
  doi     = {10.1080/00140139308967940},
}

@techreport{Waters1994,
  author      = {Waters, Thomas R. and Putz-Anderson, Vernon and Garg, Arun},
  title       = {\href{https://doi.org/10.26616/NIOSHPUB94110revised092021}{Applications Manual for the Revised {NIOSH} Lifting Equation}},
  number      = {DHHS (NIOSH) Publication No.\ 94-110 (Revised 9/2021)},
  year        = {1994},
  address     = {Cincinnati, OH}
}

@article{Karhu1977,
  author  = {Karhu, Osmo and Kansi, Pekka and Kuorinka, Iikka},
  title   = {\href{https://doi.org/10.1016/0003-6870(77)90164-8}{Correcting Working Postures in Industry: A Practical Method for Analysis}},
  journal = {Applied Ergonomics},
  volume  = {8},
  number  = {4},
  pages   = {199--201},
  year    = {1977},
  doi     = {10.1016/0003-6870(77)90164-8},
}

@article{Snook1991,
  author  = {Snook, Stover H. and Ciriello, Vincent M.},
  title   = {\href{https://doi.org/10.1080/00140139108964855}{The Design of Manual Handling Tasks: Revised Tables of Maximum Acceptable Weights and Forces}},
  journal = {Ergonomics},
  volume  = {34},
  number  = {9},
  pages   = {1197--1213},
  year    = {1991},
  doi     = {10.1080/00140139108964855},
}

@article{Generosi2022,
  author  = {Generosi, Andrea and Agostinelli, Thomas and Ceccacci, Silvia and Mengoni, Maura},
  title   = {\href{https://doi.org/10.1007/s00170-022-09880-z}{A Novel Platform to Enable the Future Human-Centered Factory}},
  journal = {The International Journal of Advanced Manufacturing Technology},
  volume  = {122},
  number  = {11},
  pages   = {4221--4233},
  year    = {2022},
  doi     = {10.1007/s00170-022-09880-z},
}

@article{Gallagher2017,
  author  = {Gallagher, Sean and Schall, Mark C.},
  title   = {\href{https://colostate.primo.exlibrisgroup.com/permalink/01COLSU_INST/5c577e/cdi_informaworld_taylorfrancis_310_1080_00140139_2016_1208848}{Musculoskeletal Disorders as a Fatigue Failure Process: Evidence, Implications and Research Needs}},
  journal = {Ergonomics},
  volume  = {60},
  number  = {2},
  pages   = {255--269},
  year    = {2017},
  doi     = {10.1080/00140139.2016.1208848},
}

@article{VinayakSen2012,
  author  = {Vinayak and Sen, Dibakar},
  title   = {\href{https://doi.org/10.1016/j.cad.2011.01.003}{A Vision Modeling Framework for {DHM} Using Geometrically Estimated {FoV}}},
  journal = {Computer-Aided Design},
  volume  = {44},
  number  = {1},
  pages   = {15--28},
  year    = {2012},
  doi     = {10.1016/j.cad.2011.01.003},
}

@misc{FreeCADWiki2024,
  author = {{FreeCAD Community}},
  title  = {{FreeCAD} Documentation: Workbench Creation},
  year   = {2024},
  url    = {https://wiki.freecad.org/Workbench_creation},
}

@misc{Assembly4,
  author = {FreeCAD},
  title  = {{FreeCAD\_Assembly4}: Assembly Workbench for {FreeCAD}},
  year   = {2024},
  url    = {https://wiki.freecad.org/Assembly4_Workbench},
}

@misc{PivyTrackers,
  author = {Graff, Joel},
  title  = {pivy\_trackers: Interactive {Coin3D} Scenegraph Node Trackers for {FreeCAD}},
  year   = {2019},
  url    = {https://github.com/joelgraff/pivy_trackers},
}

@misc{BLS2024,
  author = {{Bureau of Labor Statistics}},
  title  = {\href{https://www.bls.gov/iif/factsheets/msds.htm}{Occupational Injuries and Illnesses Resulting in Musculoskeletal Disorders (MSDs)}},
  year   = {2024},
  note   = {Injuries, Illnesses, and Fatalities (IIF) Program fact sheet, U.S. Bureau of Labor Statistics. 2021--2022 reference period. Accessed: 2026-04-28},
}

@misc{BLS2016,
  author = {{Bureau of Labor Statistics}},
  title  = {Nonfatal Occupational Injuries and Illnesses Requiring Days Away From Work, 2015},
  year   = {2016},
  url    = {https://www.bls.gov/news.release/osh2.nr0.htm},
  note   = {USDL-16-2130. Released November 10, 2016. Accessed: 2026-03-30},
}

@inproceedings{Quigley2009,
  title     = {\href{https://www.robotics.stanford.edu/~ang/papers/icraoss09-ROS.pdf}{ROS: An Open-Source Robot Operating System}},
  author    = {Quigley, Morgan and Conley, Ken and Gerkey, Brian and Faust, Josh and Foote, Tully and Leibs, Jeremy and Wheeler, Rob and Ng, Andrew Y and others},
  booktitle = {ICRA workshop on open source software},
  volume    = {3},
  number    = {3.2},
  pages     = {5},
  year      = {2009},
  organization = {Kobe},
}

@article{Paszke2019,
  title   = {\href{https://dl.acm.org/doi/10.5555/3454287.3455008}{Pytorch: An Imperative Style, High-Performance Deep Learning Library}},
  author  = {Paszke, Adam and Gross, Sam and Massa, Francisco and Lerer, Adam and Bradbury, James and Chanan, Gregory and Killeen, Trevor and Lin, Zeming and Gimelshein, Natalia and Antiga, Luca and others},
  journal = {Advances in neural information processing systems},
  volume  = {32},
  year    = {2019},
}

\end{document}